\documentclass[english,pra,reprint,showpacs]{revtex4-1}
\usepackage[T1]{fontenc}
\usepackage[latin9]{inputenc}
\usepackage{color}
\usepackage{babel}
\usepackage{verbatim}
\usepackage{amsmath}
\usepackage{graphicx}
\usepackage{xargs}[2008/03/08]
\usepackage[unicode=true,pdfusetitle,
 bookmarks=true,bookmarksnumbered=false,bookmarksopen=false,
 breaklinks=false,pdfborder={0 0 0},pdfborderstyle={},backref=false,colorlinks=true]
 {hyperref}

\makeatletter
\usepackage{babel}

\makeatother

\begin{document}

\title{Protecting weak measurements against systematic errors}

\author{Shengshi Pang$^{1,2,3}$}

\author{Jose Raul Gonzalez Alonso$^{4}$}

\author{Todd A. Brun$^{3}$}

\author{Andrew N. Jordan$^{1,2,5}$}

\affiliation{$^{1}$Department of Physics and Astronomy, University of Rochester,
Rochester, New York 14627, USA}

\affiliation{$^{2}$Center for Coherence and Quantum Optics, University of Rochester,
Rochester, New York 14627, USA}

\affiliation{$^{3}$Department of Electrical Engineering, University of Southern
California, Los Angeles, California 90089, USA}

\affiliation{$^{4}$Department of Physics and Astronomy, University of Southern
California, Los Angeles, California 90089, USA}

\affiliation{$^{5}$Institute for Quantum Studies, Chapman University, 1 University
Drive, Orange, CA 92866, USA}
\begin{abstract}
In this work, we consider the systematic error of quantum metrology
by weak measurements under decoherence. We derive the systematic error
of maximum likelihood estimation in general to the first-order approximation
of a small deviation in the probability distribution, and study the
robustness of standard weak measurement and postselected weak measurements
against systematic errors. We show that, with a large weak value,
the systematic error of a postselected weak measurement when the probe
undergoes decoherence can be significantly lower than that of a standard
weak measurement. This indicates another advantage of weak value amplification
in improving the performance of parameter estimation. We illustrate
the results by an exact numerical simulation of decoherence arising
from a bosonic mode and compare it to the first-order analytical result
we obtain.
\end{abstract}


\maketitle
\newcommandx\aw[1][usedefault, addprefix=\global, 1=]{A_{w}^{#1}}
 \global\long\def\ha{A}
 \global\long\def\hf{G}
\global\long\def\expt{{\rm expt}}
 \newcommandx\pe[2][usedefault, addprefix=\global, 1=k, 2=]{p_{#1}^{\expt}(g_{#2})}
 \newcommandx\p[3][usedefault, addprefix=\global, 1=k, 2=, 3=]{p_{#1}^{#2}(g_{#3})}
 \newcommandx\dg[1][usedefault, addprefix=\global, 1=]{\delta g_{#1}}
 \newcommandx\q[3][usedefault, addprefix=\global, 1=k, 2=, 3=]{q_{#1}^{#2}(g_{#3})}
 \global\long\def\kl{\, k=1,\cdots,d}
 \newcommandx\pg[1][usedefault, addprefix=\global, 1=]{\partial_{g}^{#1}}
 \global\long\def\rsi{\rho_{S}}
 \global\long\def\sff{|\psi_{f}\rangle}
 \newcommandx\rdi[1][usedefault, addprefix=\global, 1=]{\rho_{D}^{#1}}
 \global\long\def\rei{\rho_{E}}
 \global\long\def\i{\mathrm{i}}
 \global\long\def\e{\mathrm{e}}
 \global\long\def\re{\mathrm{Re}}
 \global\long\def\im{\mathrm{Im}}
 \global\long\def\tr{\mathrm{Tr}}
 \global\long\def\dc{\langle D|}
 \global\long\def\ed{\epsilon_{D}}
 \global\long\def\hsd{\ha\otimes\hf}
 \global\long\def\hde{H_{DE}}
 \newcommandx\csf[1][usedefault, addprefix=\global, 1=]{\langle\psi_{f}^{#1}|}
 \global\long\def\si{|\psi_{i}\rangle}
 \global\long\def\csi{\langle\psi_{i}|}
 \global\long\def\di{|D\rangle}
 \global\long\def\cdi{\langle D|}
 \global\long\def\sfi{\langle\psi_{f}|\psi_{i}\rangle}
 \global\long\def\kd{\langle k|D\rangle}
 \global\long\def\hd{\hf}
 \global\long\def\dk{\langle D|k\rangle}
 \newcommandx\hdw[1][usedefault, addprefix=\global, 1=k]{G_{w}^{(#1)}}
 \newcommandx\df[1][usedefault, addprefix=\global, 1=]{|D_{f#1}\rangle}
 \global\long\def\hm{\hat{M}}
 \global\long\def\hp{\hat{p}}
 \global\long\def\js{\langle j|\psi_{i}\rangle}
 \newcommandx\rsde[1][usedefault, addprefix=\global, 1=]{\rho_{SDE}^{#1}}
 \global\long\def\ki{|k\rangle}
 \global\long\def\cki{\langle k|}
 \global\long\def\cjk{\langle j,k|}
 \global\long\def\hdp{H_{D}^{\prime}}
 \newcommandx\hdpw[1][usedefault, addprefix=\global, 1=k]{H_{Dw}^{\prime(#1)}}
 \newcommandx\fw[1][usedefault, addprefix=\global, 1=k]{G_{w}^{(#1)}}
 \newcommandx\pp[2][usedefault, addprefix=\global, 1=, 2=]{p_{{\rm post},#1}^{#2}(g)}
 \global\long\def\sfj{|\psi_{f}^{(j)}\rangle}
 \global\long\def\asf#1{\langle\psi_{f}|#1|\psi_{f}\rangle}
 \global\long\def\asfj#1{\langle\psi_{f}^{(j)}|#1|\psi_{f}^{(j)}\rangle}
 \newcommandx\esi[1][usedefault, addprefix=\global, 1=i]{\epsilon_{S}^{#1}}
 \newcommandx\edi[1][usedefault, addprefix=\global, 1=i]{\epsilon_{D}^{#1}}
 \global\long\def\nb{b^{\dagger}b}
 \newcommandx\sdf[1][usedefault, addprefix=\global, 1=n]{|\Phi_{f}^{(#1)}\rangle}
 \newcommandx\csdf[1][usedefault, addprefix=\global, 1=n]{\langle\Phi_{f}^{(#1)}|}
 \global\long\def\hi{H_{{\rm int}}}
 \global\long\def\hip{H_{{\rm int}}^{\prime}}
 \newcommandx\sz[1][usedefault, addprefix=\global, 1=z]{\sigma_{S}^{#1}}
 \newcommandx\dz[1][usedefault, addprefix=\global, 1=z]{\sigma_{D}^{#1}}
 \global\long\def\sdz{\sz\otimes\dz}
 \global\long\def\szi{\sz[i]}
 \global\long\def\dzi{\dz[i]}
 \global\long\def\szint{(\sigma_{S}^{z})_{{\rm int}}}
 \global\long\def\dzint{(\sigma_{D}^{z})_{{\rm int}}}
 \global\long\def\vm{\overrightarrow{m}}
 \global\long\def\vn{\overrightarrow{n}}
 \global\long\def\vs{\overrightarrow{\sigma}}
 \global\long\def\mf{\mathcal{F}}
 \global\long\def\md{\mathcal{D}}
 \global\long\def\mc{\mathcal{C}}
 \global\long\def\Pet{P_{g_{0}}^{\expt}}
 \newcommandx\Pt[1][usedefault, addprefix=\global, 1=]{P_{g_{#1}}}
 \newcommandx\avg[3][usedefault, addprefix=\global, 1=, 2=]{\langle#3\rangle_{#1}^{#2}}
 \global\long\def\ml{\mathcal{L}}
\global\long\def\qt{Q_{g_{0}}}
\global\long\def\log{\ln}

\section{Introduction}

Noise is inevitable in real quantum information protocols due to the
unavoidable interactions between systems and the environment. Quantum
metrology \cite{Giovannetti2004,Giovannetti2011} is a quantum information
protocol to enhance the sensitivity of measuring physical parameters
by using quantum resources such as entanglement and squeezing. In
quantum metrology, noise usually has two detrimental effects: one
is reducing the estimation precision, and the other is biasing the
estimate. Recently, quantum metrology in open systems has been the
subject of intense study \cite{Escher2011,Demkowicz-Dobrzanski2012,Tsang2013,Kolodynski2013,Alipour2014},
and different kinds of quantum error correction techniques \cite{Lidar2013}
have been proposed to protect quantum metrology from noise, including
quantum error correcting codes \cite{Lu2015,Arrad2014,Kessler2014,Dur2014}
and dynamical decoupling \cite{Tan2013,Sekatski2015}.

In a quantum measurement, the system is generally coupled to a probe,
and the probe is measured to output the measurement results. When
the probe is measured, it is easily disturbed by external noise since
it usually must be exposed to the environment in order to be read.
This noise may come from the coupling of the probe to the environment,
the imperfection of the measurement techniques, etc. The noise can
generally have two major effects on the measurement results: decreasing
the precision and reducing the accuracy. The error correction techniques
for quantum metrology reviewed above are mostly focused on preventing
the loss of Fisher information, i.e., decrease in the measurement
precision. In this paper we consider the systematic error instead.
We study the systematic error of parameter estimation in general,
and propose using weak value amplification for parameter estimation
by weak measurement to suppress the systematic error caused by decoherence
on the probe.

Weak value amplification is an effect in postselected weak measurements
first discovered by Aharonov, Albert, and Vaidman in 1988 (AAV) \cite{Aharonov1988}.
They found that when the system is postselected to some appropriate
state in a measurement with the pointer weakly coupled to the system,
the shift of the pointer can go far beyond the eigenvalue spectrum
of the system observable in the interaction Hamiltonian. Moreover,
at the first-order of approximation the shift of the pointer is proportional
to the small interaction parameter. Thus, the measurement result can
be considered as an amplification of the interaction parameter.

Due to weak value amplification, postselected weak measurement has
been proposed to amplify small physical quantities \cite{Romito2008,Brunner2010,Feizpour2011,Li2011,Zilberberg2011,Wu2012,Dressel2013,Hayat2013,Strubi2013,Zhou2013,Pang2014b,Lyons2015,Pang2015}.
Recent state-of-the-art experimental techniques have realized the
observation of weak values \cite{Ritchie1991,Pryde2005} and applied
them to measuring small parameters in different physical systems \cite{Hosten2008,Dixon2009,Starling2009,Starling2010,Starling2010a,Pfeifer2011,Turner2011,Egan2012,Gorodetski2012,Hofmann2012,Zhou2012,Shomroni2013,Viza2013,Xu2013,Lu2014,Magana-Loaiza2014,Mirhosseini2014}.
Moreover, weak value amplification has also been found useful in quantum
state tomography \cite{Shpitalnik2008,Hofmann2010,Lundeen2011,Wu2013,Das2014,Kobayashi2014,Maccone2014}.
Reviews of the weak value technology and its applications can be found
in \cite{Kofman2012,Shikano2012,Dressel2014,Dressel2015}.

\begin{comment}
Now, let us consider the effect of weak value amplification in the
systematic error of parameter estimation in the presence of noise.
Both the system and probe may undergo noise. Noise can be introduced
by decoherence or some technical imperfections. For the noise on the
system, it will be shown that it does not influence the statistics
of the measurement outcomes in a standard weak measurement and its
effect on the average result of a postselected weak measurement also
vanishes when the probe state is chosen properly. So for both types
of weak measurements the systematic errors mainly come from the noise
on the probe.
\end{comment}

In a weak measurement, whether the system is postselected or not, noise
on the probe can introduce systematic error in the measurement results.
However, if the system is postselected, the interaction parameter
can be amplified by the weak value while the noise on the probe is
not. Thus, the weight of the probe noise in the measurement results
can be significantly suppressed by the weak value. This is the core
idea of how the systematic error may be decreased by the weak value
amplification technique.

In this paper, we study in detail the systematic error of parameter
estimation in weak measurements and the above idea of reducing the
systematic error by weak value amplification. We focus on the maximum
likelihood estimation (MLE) strategy in this work because it is the
most efficient estimator in the asymptotic limit. We  first obtain
a general result for the systematic error of MLE in the first-order
of a small deviation in the probability distribution. Then, we apply
it to the standard weak measurement and the postselected weak measurement,
and prove that the systematic error of the weak measurement can indeed
be reduced by postselecting the system with a large weak value. We
illustrate the result by a numerical simulation of a simple example
with a qubit system, a qubit probe, and a single bosonic mode thermal
bath.

It is worth mentioning that recently there has been a controversy
over the precision of weak value amplification \cite{Starling2009,Starling2010a,Feizpour2011,Zhu2011,Dressel2013,Knee2013,Tanaka2013,Combes2014,Ferrie2014,Jordan2014,Knee2014,Pang2014b,Torres2016,Lyons2015,Pang2015a,Pang2015,Zhang2015a}.
Since the Fisher information is proportional to the amount of the
data and the postselection of the system discards a large portion
of the measurement results, then the precision of the measurement
may be lowered by the postselection. However, it was shown that the
loss of Fisher information can be negligible when the initial state
and the postselected state of the system are chosen properly \cite{Jordan2014,Pang2014b,Pang2015,Viza2015a,Alves2015},
and that the signal-to-noise ratio (SNR) of postselected weak measurements
can be made much higher than that of standard weak measurements by
utilizing squeezed states for the probe \cite{Pang2015a}.

In contrast to the Fisher information, the systematic error of estimation
does not scale with the amount of data. Therefore, if one can select
the measurement results that deviate less from the true value of the
parameter, and discard the unselected events, the accuracy of the
estimate can be enhanced by proper postselection without suffering
from the low postselection probability. This is how the systematic
error differs from the Fisher information, and it is how we can avoid
the problem of a low postselection probability that is an issue when
using the Fisher information.

The structure of the paper is as follows. First, we review the weak
value formalism for postselected weak measurements in Sec. \ref{sec:weak value formalism}.
Then, in Sec. \ref{sec:Systematic-error-of} we study the systematic
error of maximum likelihood estimation in the first-order approximation
when the probability distribution deviates slightly from the ideal.
The following section \ref{sec:Weak-measurement-with} is devoted
to a detailed investigation of the systematic errors of standard weak
measurements and postselected weak measurements and we show the advantage
of postselecting the system in protecting the measurement accuracy
against decoherence. A simple qubit example is used in Sec. \ref{sec:Numerical-example}
to explicitly illustrate the analytical result by a numerical computation.

\section{Review of weak value theory\label{sec:weak value formalism}}

The effects of postselected weak measurements are usually characterized
by weak values. (See \cite{Dressel2014} for a review of weak values.)
In this section, we briefly review the weak value formalism for postselected
weak measurement as a foundation for the discussion in the following
sections.

In a quantum measurement, a measuring device coupled to the system
is typically modeled by an interaction Hamiltonian between the system
and the measuring device that can be written as
\begin{equation}
H_{I}=g\ha\otimes\hf\delta(t-t_{0}),\label{eq:12-1}
\end{equation}
where $\ha$ and $\hf$ are observables of the system and the pointer,
respectively, and $g$ characterizes the strength of the interaction.
After the system and the measuring device are coupled, the device
is then measured and outputs the measurement result.

Suppose the initial state of the system is $\si$, while the initial
state of the pointer is $\di$, then the system and the pointer are
coupled by the interaction, and evolve to an entangled state
\begin{equation}
|\Phi\rangle=\exp(-\i g\ha\otimes\hf)\si\di.
\end{equation}
The evolved state $|\Phi\rangle$ can be written as
\begin{equation}
|\Phi\rangle=\sum_{k}c_{k}|a_{k}\rangle\exp(-\i ga_{k}\hf)\di,\label{eq:19}
\end{equation}
where $a_{k},\,|a_{k}\rangle$ are eigenvalues and eigenstates of
$\ha$, and $c_{k}$ are the expansion coefficients of $\si$ in the
basis $\{|a_{k}\rangle\}$.

In a projective quantum measurement, the coupling between the system
and the pointer is usually sufficiently strong, so that the overlaps
between different $\exp(-\i ga_{k}\hf)\di$ are very small. In this
case, the $\exp(-\i ga_{k}\hf)\di$ can be distinguished with a low
error probability, and the measurement to distinguish different $\exp(-\i ga_{k}\hf)\di$
will collapse the system to a state close to an eigenstate of $\ha$.
This is what the theory of standard projective quantum measurement
tells us: the results of a projective quantum measurement are the
eigenvalues of the observable that is measured, and the system will
collapse to an eigenstate of that observable. On the contrary, in
a weak measurement, the coupling between the system and the pointer
is usually very weak, and different $\exp(-\i ga_{k}\hf)\di$ may
substantially overlap.

The invention of AAV in weak measurements is to introduce postselection
to the system which is weakly coupled to the measuring device. This
small change gives dramatically different results from the standard
projective quantum measurements. If the system is postselected to
the state $\sff$ after it is coupled to the measuring device, then
the measuring device collapses to
\begin{equation}
|D_{f}\rangle=\csf\exp(-\i g\ha\otimes\hf)\si\di.
\end{equation}
In a weak measurement $g$ is usually very small therefore $\df$
can be approximated by
\begin{equation}
\begin{aligned}\df & \approx\csf(1-\i g\ha\otimes\hf)\si\di\\
 & =\sfi(1-\i g\aw\hf)\di,
\end{aligned}
\label{eq:11-1}
\end{equation}
where $\aw$ is defined as a weak value,
\begin{equation}
\aw=\frac{\csf\ha\si}{\sfi}.\label{eq:8-1}
\end{equation}

If $g\aw\ll1$ is satisfied, $\df$ can be rewritten as
\begin{equation}
\df\approx\exp(-\i g\aw\hf)\di.
\end{equation}
Therefore, when the system is postselected after the weak interaction
with the measuring device, the measuring device is approximately rotated
by $\exp(-\i g\aw\hf)$. This is in sharp contrast to the standard
projective measurement, because $\aw$ can be much larger than $1$
when $|\sfi|\ll1$, and the rotation of the measuring device can be
much larger than the eigenvalues of $\ha$ in the postselected case.

Note that $\aw$ can be complex, and in this case $\exp(-\i g\aw\hf)$
is not just a simple translation operator. In fact, it can be decomposed
as the product of a translation operator (corresponding to the real
part of $\aw$) and a state reduction operator (corresponding to the
imaginary part of $\aw$). Jozsa gave a very detailed study of complex
weak values in Ref. \cite{Jozsa2007} and analyzed the role of the
real and imaginary parts of the weak value. He showed that, if the
pointer observable $\hf$ is the momentum $\hat{p}$, then the shifts
in the average position and momentum of the pointer are, respectively,
\begin{equation}
\begin{aligned}\langle\Delta\hat{q}\rangle= & g\mathrm{Re}\aw+gm\mathrm{Im}\aw\frac{\mathrm{d}}{\mathrm{d}t}\mathrm{Var}(\hat{q}),\\
\langle\Delta\hat{p}\rangle= & 2g\mathrm{Im}\aw\mathrm{Var}(\hat{p}),
\end{aligned}
\label{eq:33}
\end{equation}
where $\hat{q}$ and $\hat{p}$ are the position and momentum operators
of the pointer, and $m$ is the mass of the pointer.

That result can be cast in a more general form. Suppose we measure
an observable $\hm$ on the pointer after postselecting the system.
The average shift of the pointer is
\begin{equation}
\langle\Delta\hm\rangle_{f}=\frac{\langle D_{f}|\hm|D_{f}\rangle}{\langle D_{f}|D_{f}\rangle}-\langle\hm\rangle_{\di}.\label{eq:6-1}
\end{equation}
From Eq. (\ref{eq:11-1}), one can get
\begin{equation}
\begin{aligned}\langle D_{f}|\hm|D_{f}\rangle\approx & |\sfi|^{2}[\langle\hm\rangle_{\di}+\i g\re\aw\langle[\hf,\hm]\rangle_{\di}\\
 & +g\im\aw\langle\{\hf,\hm\}\rangle_{\di}],\\
\langle D_{f}|D_{f}\rangle\approx & |\sfi|^{2}[1+2g\im\aw\langle\hf\rangle_{\di}],
\end{aligned}
\label{eq:3-1}
\end{equation}
thus,
\begin{equation}
\begin{aligned}\langle\Delta\hm\rangle_{f}\approx & g\im\aw(\langle\{\hf,\hm\}\rangle_{\di}-2\langle\hf\rangle_{\di}\langle\hm\rangle_{\di})\\
 & +\i g\re\aw\langle[\hf,\hm]\rangle_{\di},
\end{aligned}
\label{eq:7-1}
\end{equation}
which is similar to the result in \cite{Dressel2012}. Note that if
we plug $\hf=\hat{p}$ and $\hm=\hat{q},\,\hat{p}$ into (\ref{eq:7-1}),
the result in (\ref{eq:33}) can be immediately recovered.

Eqs. (\ref{eq:33}) and (\ref{eq:7-1}) imply that the shift of the
pointer is roughly proportional to the weak value $\aw$ when $g\ll1$.
Since $\aw$ can be much larger than $1$ when $\sfi\ll1$, the shift
of the pointer can be treated as an amplification of $g$ by the weak
value $\aw$. This is the origin of the amplification effect in postselected
weak measurements. This amplification effect has been widely used
in experiments to measure small parameters, as reviewed in the introduction.

\section{Systematic error of maximum likelihood estimation\label{sec:Systematic-error-of}}

In maximal likelihood estimation (MLE), suppose we want to estimate
an unknown parameter $g$ from a $g$-dependent probability distribution
$\Pt:\,\p,\kl.$ However, due to the interaction with a noisy environment,
the real probability distribution observed in experiments is $\Pet:\,\pe[k],\kl$,
which slightly deviates from $\p$. Then, the estimate of $g$ will
generally deviate from the \emph{true value} $g_{0}$, i.e., a systematic
error may occur in this case. In this section, we derive a general
first-order solution to the systematic error for such a noisy MLE.

Suppose we observe the result $k$ a total of $N_{k}$ times in an
experiment, and the measurement results are uncorrelated. The spirit
of MLE is finding the most likely $g$ (the parameter to estimate)
conditioned on the observation results as the estimate for $g$. In
a mathematical language, it means to maximize the following likelihood
function over $g$: $\ml=\prod_{k}\p[][N_{k}]$, or alternatively
its logarithm $\log\ml=\sum_{k}N_{k}\log\p$.

When the total number of measurement results $N=\sum_{k}N_{k}$ is
very large, $N_{k}\approx N\pe[][0]$ in average, so the MLE leads
to the following equation with respect to $g$:
\begin{equation}
\pg\log\ml\approx N\sum_{k}\pe[][0]\frac{\partial_{g}\p}{\p}=0.\label{eq:1}
\end{equation}
This equation usually produces multiple solutions for $g$, and we
need to find the one that has the largest likelihood. It should be
noted that the derivative with respect to $g$ is always performed
on $\p$ in Eq. \eqref{eq:1}, because $\pe[][0]$ is the average
frequency that the $k$-th result will be observed in an experiment
and $g_{0}$ is the true value of $g$ (which is a constant).

When $\pe=\p,\kl,$ it is obvious that $g=g_{0}$ is the solution
to (\ref{eq:1}) since $\sum_{k}\p=1$, which implies that the MLE
is an unbiased estimation strategy in this case. However if $\pe\neq\p,$
then generally $g\neq g_{0}$, which leads to a systematic error.
(Note that the systematic error cannot be eliminated by repetition
of the measurements.) If we write $g=g_{0}+\dg$, then the question
is how large $\dg$ is in terms of the deviation of $\p$ from $\pe$.

Suppose $\pe=\p+\q,\kl,$ where $|\q|\ll1$, and $\sum_{k}\q=0$.
In this case, Eq. (\ref{eq:1}) becomes
\begin{equation}
\sum_{k}(\p[][][0]+\q[][][0])\frac{\pg\p}{\p}=0.
\end{equation}
If we expand $\frac{\pg\p}{\p}$ to the first-order of $\dg$, then
\begin{equation}
\begin{aligned} & \sum_{k}(\p[][][0]+\q[][][0])\bigg(\left.\frac{\pg\p}{\p}\right|_{g=g_{0}}\\
 & +\dg\left.\frac{\p\pg[2]\p-(\pg\p)^{2}}{\p[][2][0]}\right|_{g=g_{0}}\bigg)=0.
\end{aligned}
\end{equation}
Up to the first-order of $|\q|$, we get
\begin{equation}
\dg\approx\frac{\pg\mc(\qt||\Pt)|_{g=g_{0}}}{\mf(\Pt[0])},\label{eq:2}
\end{equation}
where $\mc(\qt||\Pt)$ is
\begin{equation}
\mc(\qt||\Pt)=\sum_{k}\q[][][0]\log\p,
\end{equation}
and $\mf(\Pt)$ is the Fisher information of the probability distribution
$\Pt$ at $g=g_{0}$,
\begin{equation}
\mf(\Pt[0])=\sum_{k}\left.\frac{(\pg\p)^{2}}{\p}\right|_{g=g_{0}}.
\end{equation}

Since $\q=\pe-\p$ and the derivative is with respect to $g$ only,
$\pg\mc(\qt||\Pt)$ can be written as
\begin{equation}
\begin{aligned} & \partial_{g}\mc(\qt||\Pt)\\
 & =\pg\sum_{k}\pe[][0](\log\p-\log\pe[][0])\\
 & -\pg\sum_{k}\p[][][0]\log\p.
\end{aligned}
\label{eq:23}
\end{equation}
At $g=g_{0}$,
\begin{equation}
\sum_{k}\p[][][0]\pg\log\p|_{g=g_{0}}=0,
\end{equation}
so only the first term on the right side of Eq. (\ref{eq:23}) is
nonzero. Thus,
\begin{equation}
\partial_{g}\mc(\qt||\Pt)|_{g=g_{0}}=-\pg\md(\Pet||\Pt)|_{g=g_{0}},
\end{equation}
where $\md(\Pet||\Pt)$ is the relative entropy between $\Pet$ and
$\Pt$,
\begin{equation}
\md(\Pet||\Pt)=\sum_{k}\pe[][0]\log\frac{\pe[][0]}{\p}.
\end{equation}
Therefore, the first-order systematic error $\dg$ given in (\ref{eq:2})
can be finally written as
\begin{equation}
\dg=-\frac{\pg\md(\Pet||\Pt)|_{g=g_{0}}}{\mf(\Pt[0])},\label{eq:sys-error}
\end{equation}
the ratio of the derivative of relative entropy to the Fisher information.

Before concluding this section, we want to mention a substantial difference
between the systematic error and the variance of an estimator. The
Cramér-Rao bound \cite{Cramer1946} tells us that
\begin{equation}
\langle\dg^{2}\rangle\geq\frac{1}{N\mf_{g}}+\langle\dg\rangle^{2}.\label{eq:camer-rao-bound}
\end{equation}
The first term is the inverse of the Fisher information, which is
the lower bound of the variance of the estimator. It can be seen that
the variance scales as $N^{-1}$, the standard quantum limit. In contrast,
the second term, i.e., the systematic error, does not depend on $N$,
which implies it cannot be changed by the number of measurements.

This observation has two important implications. The first one is
that systematic errors cannot be reduced simply by increasing the
number of measurements, as random noise is usually treated, so when
$N\rightarrow\infty$, the random errors will approach zero but the
systematic errors will remain finite, thus it calls for new methods
to overcome the systematic errors. The other is that if weak value
amplification can reduce the systematic error (as we will show later),
the low postselection probability will not affect it, since the systematic
error is not proportional to the size of the data. This is the foundation
of weak value amplification in suppressing systematic errors in weak
measurements.

\section{Weak measurements with decoherence\label{sec:Weak-measurement-with}}

In this section, we study the systematic error of weak measurements
with or without postselection in the presence of decoherence and show
the effect of large weak values in suppressing the systematic errors.

We first give some general analysis about the systematic error in
a standard or a postselected weak measurement. From Eq. (\ref{eq:sys-error}),
we can see that the systematic error of the MLE is the reciprocal
of the Fisher information. Since the Fisher information can be amplified
by the order of $|\aw|^{2}$ when the system is postselected, and
it cannot be reduced by the postselection probability since it is
not dependent on the size of data according to (\ref{eq:sys-error}),
then a large weak value may reduce the systematic error. Note that
$\pg\md(\Pet||\Pt)$ may also be amplified by the weak value because
the probability distribution of the measurement results on the probe
is approximately shifted by $g\aw$ according to (\ref{eq:7-1}),
but the order of the amplification factor is $|\aw|$. So the net
effect of weak value amplification is to reduce the systematic error
by the order of $|\aw|$.

Below, we study the systematic error of weak measurements and the
effect of weak value amplification on reducing it in detail. Throughout
this paper, we let $\hbar=1$ for simplicity.

When the pointer undergoes decoherence, a typical interaction Hamiltonian
is
\begin{equation}
H_{I}=g\hsd\delta(t-t_{0})+\ed\hde,\,g,\ed\ll1,
\end{equation}
where the subscripts $S$, $D$, $E$ represent the system, the probe
and the environment respectively. For simplicity, we assume that $g,\ed\ll1$
and that $g^{2}\ll\ed$ so that the second order of $g$ can be neglected.
Suppose the initial states of the system and the probe are $\rsi,\,\rdi,$
and the initial state of the environment is $\rei$. Then, after time
$t$, the joint state of the system, probe and environment is
\begin{equation}
\rsde[\expt](t)=\exp(-\i H_{I}t)\rsi\otimes\rdi\otimes\rei\exp(\i H_{I}t).
\end{equation}
When the interaction time is very short, i.e., $t\ll1,$ we have
\begin{equation}
\begin{aligned}\rsde[\expt](t)\approx & \rsi\otimes\rdi\otimes\rei-\i[g\hsd\\
 & +\ed t\hde,\,\rsi\otimes\rdi\otimes\rei].
\end{aligned}
\end{equation}

If there is no decoherence on the probe, $\ed=0$, the joint state
of the system, probe and environment can be reduced to
\begin{equation}
\rsde(t)=(\rsi\otimes\rdi-\i g[\hsd,\,\rsi\otimes\rdi])\otimes\rei.
\end{equation}
So, the joint state $\rsde[\exp](t)$ in the presence of decoherence
can be rewritten as
\begin{equation}
\rsde[\expt](t)=\rsde(t)-\i t[\ed\hde,\,\rsi\otimes\rdi\otimes\rei].
\end{equation}

\subsection{Standard weak measurement}

When there is no postselection on the system in the weak measurement,
the post-interaction pointer state in the absence of decoherence is
\begin{equation}
\begin{aligned}\rdi(t) & \approx\rdi-\i g\tr_{S}[\hsd,\,\rsi\otimes\rdi]\\
 & \approx\rdi-\i g\avg[i]{\ha}[\hf,\,\rdi],
\end{aligned}
\label{eq:std_rhod}
\end{equation}
where
\begin{equation}
\avg[i]{\ha}=\tr_{S}(\ha\rsi).
\end{equation}
And the post-interaction probe state in the presence of decoherence
is
\begin{equation}
\begin{aligned}\rdi[\expt](t) & \approx\rdi(t)-\i t\tr_{SE}[\ed\hde,\,\rsi\otimes\rdi\otimes\rei]\\
 & =\rdi(t)-\i t\ed[\hdp,\,\rdi],
\end{aligned}
\end{equation}
where
\begin{equation}
\hdp=\tr_{E}(\hde\rei).\label{eq:hdp}
\end{equation}

Now, if we measure an orthonormal basis $\{\ki\}$ on the post-interaction
probe state, the probability distribution in the decoherence-free
case is
\begin{equation}
\p=\cki\rdi\ki-\i g\avg[i]{\ha}\cki[\hf,\rdi]\ki,
\end{equation}
and the probability distribution for the case with decoherence is
\begin{equation}
\pe=\p+\q,
\end{equation}
where
\begin{equation}
\q=-\i\ed t\cki[\hdp,\rdi]\ki.
\end{equation}

Then, since $g\ll1,$ we have
\begin{equation}
\begin{aligned}\pg\md(\Pet||\Pt)|_{g=g_{0}} & \approx\ed t\avg[i]{\ha}\times\\
 & \sum_{k}\frac{\cki[\hdp,\rdi]\ki\cki[\hf,\rdi]\ki}{\cki\rdi\ki},\\
\mf(\Pt[0]) & \approx\avg[i][2]{\ha}\sum_{k}\frac{|\cki[\hd,\rdi]\ki|^{2}}{\cki\rdi\ki}.
\end{aligned}
\label{eq:5}
\end{equation}
If we define the following weak values for $\hd$ and $\hdp$
\begin{equation}
\hdw=\frac{\cki\hd\rdi\ki}{\cki\rdi\ki},\,\hdpw=\frac{\cki\hdp\rdi\ki}{\cki\rdi\ki},\label{eq:8}
\end{equation}
then \eqref{eq:5} can be simplified to
\begin{equation}
\begin{aligned}\pg\md(\Pet||\Pt)|_{g=g_{0}} & \approx-4\ed t\avg[i]{\ha}\times\\
 & \sum_{k}\cki\rdi\ki\im\hdpw\im\hdw[k],\\
\mf(\Pt[0]) & \approx4\avg[i][2]{\ha}\sum_{k}\cki\rdi\ki\im^{2}\hdw[k].
\end{aligned}
\end{equation}

Therefore, according to Eq. (\ref{eq:sys-error}), the systematic
error $\dg[n]$ of the standard weak measurement which has no postselection
on the system is approximately
\begin{equation}
\dg[n]\approx\frac{\ed t{\displaystyle \sum_{k}\cki\rdi\ki\im\hdpw\im\hdw[k]}}{\avg[i]{\ha}{\displaystyle \sum_{k}\cki\rdi\ki\im^{2}\hdw[k]}}.\label{eq:6}
\end{equation}

From the above result, we can see that generally the initial state
of the system should not be chosen such that $\avg[i]{\ha}\ll1$,
otherwise, the Fisher information of the measurement in Eq. (\ref{eq:5})
will be very small, implying a bad estimation precision, and the systematic
error in Eq. (\ref{eq:6}) will be very large, implying a large deviation
of the measurement result from the true value of the parameter.

\subsection{Postselected weak measurement}

In a postselected weak measurement, the system is postselected to
some specific state $\sff$ after it is coupled to the probe. So the
pointer state after the postselection of the system in the absence
of decoherence is
\begin{equation}
\begin{aligned}\rdi(t)\approx & \asf{\rsi}\rdi-\i g\asf{[\hsd,\rsi\otimes\rdi]}\\
= & \asf{\rsi}(\rdi-\i g\re\aw{}[\hf,\,\rdi]\\
 & +g\im\aw\{\hf,\,\rdi\}).
\end{aligned}
\label{eq:14}
\end{equation}
\begin{comment}
After normalization, $\rdi(t)$ becomes
\begin{equation}
\begin{aligned}\rdi(t) & \approx\rdi-\frac{\i g}{\asf{\rsi}}(\asf{[\hsd,\rsi\otimes\rdi]}\\
 & -\asf{[\hs,\rsi]}\rdi),
\end{aligned}
\end{equation}
where $\hs=\tr_{D}(\hsd\rdi)$, and the second term between the parentheses
in the second line comes from the normalization. And in the presence
of decoherence, the post-interaction probe state is
\begin{equation}
\begin{aligned}\rdi[\exp](t) & \approx\asf{\rsi}\rdi-\i g\asf{[\hsd,\rsi\otimes\rdi]}\\
 & -\i\es t\asf{[\hsp,\rsi]}\rdi\\
 & -\i\ed t\asf{\rsi}[\hdp,\rdi]\\
 & \propto\rdi-\i g\frac{\asf{[\hsd,\rsi\otimes\rdi]}}{\asf{\rsi}}\\
 & -\i\es t\frac{\asf{[\hsp,\rsi]}}{\asf{\rsi}}\rdi-\i\ed t[\hdp,\rdi],
\end{aligned}
\label{eq:7}
\end{equation}
\end{comment}
In the presence of decoherence on the probe, the post-interaction
probe state formed by tracing out the environment is
\begin{equation}
\begin{aligned}\rdi[\expt](t)\approx & \rdi(t)-\i\ed t\asf{\rsi}[\hdp,\rdi],\\
\propto & \rdi-\i g\re\aw{}[\hf,\,\rdi]\\
 & +g\im\aw\{\hf,\,\rdi\}-\i\ed t[\hdp,\rdi],
\end{aligned}
\end{equation}
where $\hdp$ is defined in Eq. (\ref{eq:hdp}). %
\begin{comment}
After normalization, $\rdi[\exp](t)$ becomes (with the first-order
approximation of $g,\,\es,\,\ed$)
\begin{equation}
\rdi[\exp](t)\approx\rdi(t)-\i\ed t[\hdp,\rdi],
\end{equation}
where the system decoherence has been canceled, and $\tr_{D}[\hdp,\rdi]=0$
has been used.
\end{comment}

Now, if we measure along an orthonormal basis $\{|k\rangle\}$ on
the post-interaction pointer state, the probability distribution of
the measurement results in the decoherence-free case is
\begin{equation}
\begin{aligned}\p= & \langle k|\rdi|k\rangle-\i g\re\aw\langle k|[\hf,\,\rdi]|k\rangle\\
 & +g\im\aw\langle k|\{\hf,\,\rdi\}|k\rangle.
\end{aligned}
\end{equation}
If we define the following weak value for $\hf$,
\begin{equation}
\fw=\frac{\cki\hf\rdi\ki}{\langle k|\rdi|k\rangle},
\end{equation}
\begin{comment}
which satisfies $\hs=\sum_{k}\langle k|\rdi|k\rangle\hsk$, and define
weak values
\begin{equation}
\begin{aligned}\hskw & =\frac{\asf{\hsk\rsi}}{\asf{\rsi}},\\
\hswo & =\frac{\asf{[\hs,\rsi]}}{\asf{\rsi}},
\end{aligned}
\end{equation}
then Eq. (\ref{eq:3}) can be rewritten as
\begin{equation}
\p=\langle k|\rdi|k\rangle[1+2g(\im\hskw-\im\hswo)].\label{eq:3}
\end{equation}
\end{comment}
then,
\begin{equation}
\begin{aligned}\p & =\langle k|\rdi|k\rangle[1+2g(\re\aw\im\fw+\im\aw\re\fw)]\\
 & =\langle k|\rdi|k\rangle[1+2g\im(\aw\fw)].
\end{aligned}
\end{equation}

Also, the probability distribution of the measurement results in the
presence of decoherence is
\begin{equation}
\pe=\p+\q,\label{eq:15-1}
\end{equation}
and
\begin{equation}
\q=2\langle k|\rdi|k\rangle t\ed\im\hdpw,
\end{equation}
where $\hdpw$ is defined as
\begin{equation}
\hdpw=\frac{\cki\hdp\rdi\ki}{\langle k|\rdi|k\rangle}.
\end{equation}
\begin{comment}
In the weak interaction limit $g\ll1$, it can be derived that
\begin{equation}
\begin{aligned} & \partial_{g}H_{p}(\{\q[][][0]\},\{\p\})|_{g=g_{0}}\\
 & \approx2\ed t\sum_{k}\langle k|\rdi|k\rangle\im\hdpw(\im\hskw-\im\hswo)\\
 & =-4\ed t\sum_{k}\langle k|\rdi|k\rangle\im\hdpw\im\hskw,\\
 & F_{p}(\{\p[][][0]\})\\
 & \approx4\sum_{k}\langle k|\rdi|k\rangle(\im\hskw-\im\hswo)^{2}\\
 & =4(\sum_{k}\langle k|\rdi|k\rangle\im^{2}\hskw-\im^{2}\hswo),
\end{aligned}
\end{equation}
where we have used $\sum_{k}\langle k|\rdi|k\rangle\im\hdpw=0$ and
$\sum_{k}\langle k|\rdi|k\rangle\im\hskw=\im\hswo$. The latter equation
is because $\sum_{k}\langle k|\rdi|k\rangle\hsk=\hs$.
\end{comment}

In the weak interaction limit $g\ll1$, we can see that
\begin{equation}
\begin{aligned}\partial_{g}\md(\Pet||\Pt)|_{g=g_{0}} & \approx4t\sum_{k}\langle k|\rdi|k\rangle\ed\im\hdpw\im(\aw\fw)\\
\mf(\Pt[0]) & \approx4\sum_{k}\langle k|\rdi|k\rangle\im^{2}(\aw\fw).
\end{aligned}
\end{equation}
Therefore, the systematic error $\dg[p]$ when the system is postselected
to $\sff$ is
\begin{equation}
\dg[p]\approx\frac{{\displaystyle \ed t\sum_{k}\langle k|\rdi|k\rangle\im\hdpw\im(\aw\fw)}}{{\displaystyle \sum_{k}\langle k|\rdi|k\rangle\im^{2}(\aw\fw)}}.\label{eq:30}
\end{equation}

When the postselection probability $\asf{\rsi}\ll1,$ the weak value
$\aw$ of system observable $\ha$, is of order $\asf{\rsi}^{-\frac{1}{2}}$,
which can be very large. And $\dg[p]$ is proportional to the inverse
of the system weak value $\aw$. (Note that there is no average over
$\aw$.) Therefore, this implies that the systematic error due to
the decoherence can be suppressed by a large weak value, when the
system is postselected to a state that is almost orthogonal to its
initial state.

A notable point here is that if we know $\hdp$ exactly, there may
be an even simpler (and perhaps more efficient) way to decrease the
systematic error. That is, we can simply choose proper initial state
and measurement basis for the probe such that $\partial_{g}\md(\Pet||\Pt)$
{[}i.e., the numerator of Eq. \eqref{eq:6}{]} is close to zero, in
which case the systematic error becomes extremely small (it is generally
still nonzero because only the first-order terms of $g$ and $\ed$
are considered in the above results). For example, provided that we
have detailed knowledge about $\hdp$, we can choose a basis $\{|k\rangle\}$
for the measurement on the probe such that all $\hdpw$ are real,
thus $\im\hdpw=0$ for all $k$, and $\dg[n]$ (and $\dg[p]$) would
be approximately zero. This is a simpler way to suppress the systematic
error, and postselection of the system is not so necessary in this
case.

However, in practice, the interaction between the probe and the environment
can be very complex, and one generally does not have complete information
about the decoherence, so it is usually not practical to decrease
the systematic error in this simpler way. In contrast, the method
proposed above, which is based on the weak value amplification, only
requires a large weak value $\mbox{\ensuremath{\aw}}$, regardless
of the details of the decoherence, so this method provides a universal
approach for reducing the systematic error in weak measurements.

\section{Numerical example\label{sec:Numerical-example}}

To illustrate the results derived in the previous section, we consider
a simple qubit dephasing model and use a numerical computation to
showcase the effects of postselection in the estimation of the system
and probe coupling while varying different parameters of the model.

Suppose the total Hamiltonian for the system, probe, and environment
is
\begin{equation}
H=g\sdz\delta(t-t_{0})+\ed\dz[y]\otimes\nb,\label{eq:9}
\end{equation}
where $g,\,\ed\ll1$ and the environment is assumed to have a single
bosonic mode $b$. In Eq. \eqref{eq:9}, we only considered the coupling
between the $y$-components of the probe qubit and the environment
for simplicity. Generally, there can also be coupling between $x$,
and $z$-components of the qubit and the environment.

Suppose the system and the probe are initially in states $\si$ and
$\di$, and that the environment is initially in the thermal equilibrium
state $\rei$,
\begin{equation}
\rho_{E}=\frac{1}{Z}\exp(-\beta\nb),\,\beta=\frac{\omega}{kT},
\end{equation}
where $\omega$ is the frequency of the single mode oscillator of
the environment, $k$ is the Boltzmann constant, and $T$ is the temperature
of the environment. The partition function $Z$ is given by
\begin{equation}
Z=\tr\exp(-\beta\nb)=\frac{1}{1-\exp(-\beta)}.
\end{equation}

After a short time $t$, the joint state of the system and probe evolves
to
\begin{equation}
\rho_{SD}=\frac{1}{Z}\sum_{n}\e^{-\i\beta n}\sdf[n]\csdf,\label{eq:18}
\end{equation}
where $\sdf$ is
\begin{equation}
\sdf=\exp[-\i(g\sdz+tn\ed\dz[y])]\si\di.\label{eq:20}
\end{equation}

In our numerical simulation, we chose $\si=|+\rangle$ and $\di=|+\rangle$
as the initial states for the system and the probe, respectively.
The postselected state of the system is
\begin{equation}
\sff=\exp\left(-\i\delta\sigma_{S}^{y}\right)|-\rangle,\,\delta\ll1.\label{eq:10}
\end{equation}
Hence, the weak value of $\sigma_{z}$ is
\begin{equation}
(\sigma_{S}^{z})_{w}=\frac{\csf\sigma_{S}^{z}\si}{\sfi}=\cot\delta,\label{eq:11}
\end{equation}
which is approximately $1/\delta$ when $\delta\ll1$. The measurement
basis that we choose on the probe is
\begin{equation}
|k'\rangle=\e^{-\i\theta\sigma^{x}}|k\rangle,\;k=0,\,1,
\end{equation}
where $\theta$ is a parameter to adjust.%
\begin{comment}
Then the post-interaction state of the system and the probe is
\begin{equation}
\begin{aligned}\di[f][SD] & =\tr_{E}(\langle\e^{-\i H_{I}t}|+\rangle_{S}|+\rangle_{D}\rho_{E}\langle+|_{D}\langle+|_{S})\e^{\i H_{I}t}\\
 & =\sum_{n}\exp(-\i g\sigma_{z}^{S}\otimes\sigma_{z}^{D}-\i\es tn\sigma_{y}^{S}-\i\ed tn\sigma_{y}^{D})|+\rangle_{S}|+\rangle_{D}\\
 & \langle+|_{D}\langle+|_{S}\exp(-\i g\sigma_{z}^{S}\otimes\sigma_{z}^{D}-\i\es tn\sigma_{y}^{S}-\i\ed tn\sigma_{y}^{D}).
\end{aligned}
\end{equation}
If the system is postselected to the state $\sf$ \eqref{eq:10},
the probe state collapses to
\end{comment}

Below we plot the numerical results for this example varying the values
of the different parameters of interest. Unless otherwise noted, we
fix $\beta=1.0,\,\theta=\frac{\pi}{8},\,\delta=1.0\times10^{-3},\,g=1.0\times10^{-5}$,
and $\edi=1.0\times10^{-5}$. In all plots, we show the ratio of the
systematic uncertainty with and without postselection on the system,
$\frac{\dg[p]}{\dg[n]}$.

\begin{figure}[!htbp]
\includegraphics[width=0.95\linewidth]{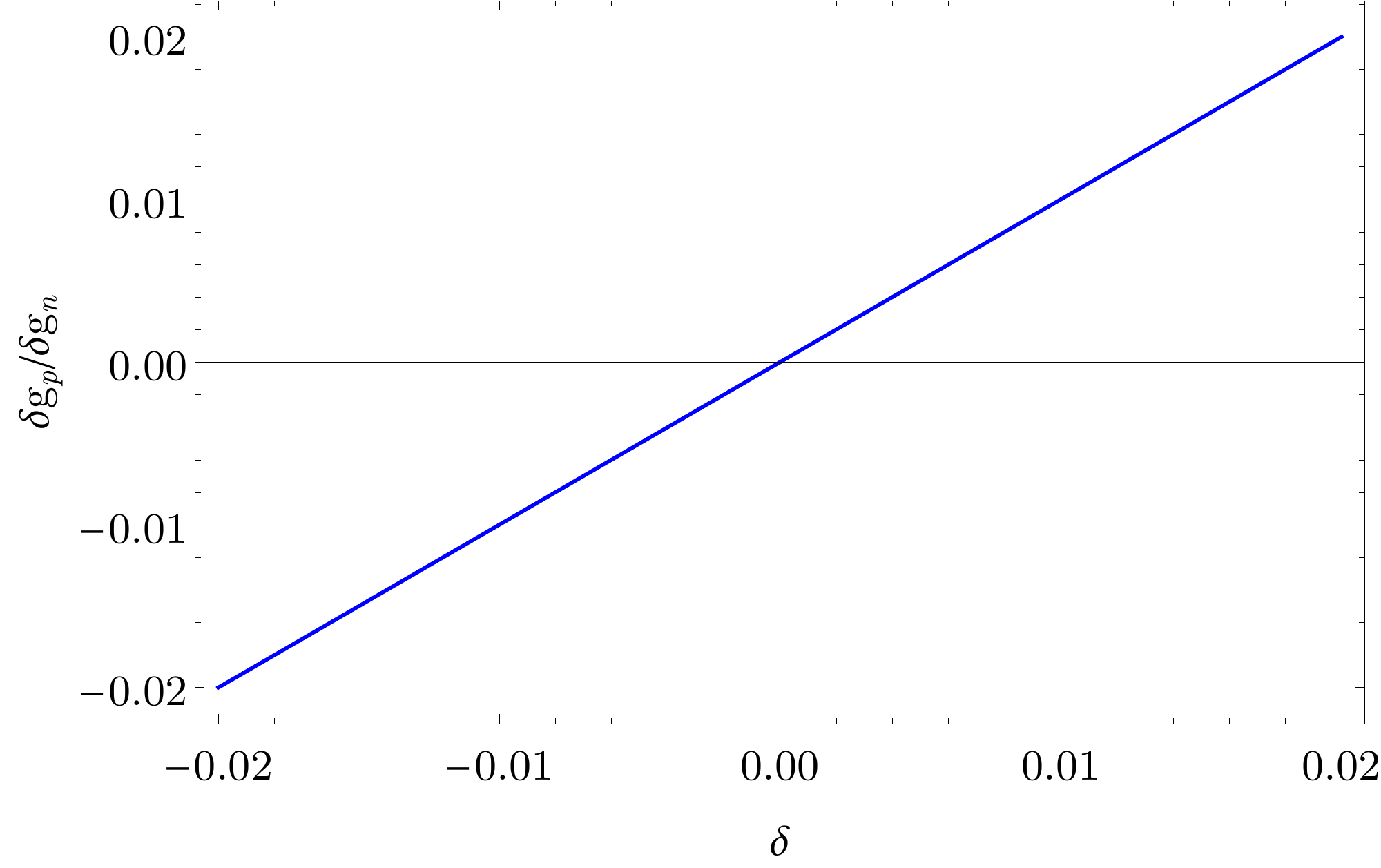}
\caption{Ratio between the systematic error with and without postselection
of the system for postselected states with varying $\delta$.}
\label{fig:lp1dOverlp3d}
\end{figure}

Fig. \ref{fig:lp1dOverlp3d} shows the ratio $\frac{\dg[p]}{\dg[n]}$
with different choices of the postselected state with varying $\delta$.

As we can see in Fig. \ref{fig:lp1dOverlp3d}, when comparing the
systematic error of the estimation with and without postselection,
we find that the postselection will suppress the systematic error
of estimation by the order of $\delta$. This is expected because
we see from Eqs. \eqref{eq:6} and \eqref{eq:30} that when all the
parameters but $\delta$ are kept fixed, the systematic error without
postselection does not depend on $\delta$, while for the postselection
case, the systematic error is approximately proportional to $\delta$
when $\left|\delta\right|\ll1$.

\begin{figure}[!htbp]
\includegraphics[width=0.95\linewidth]{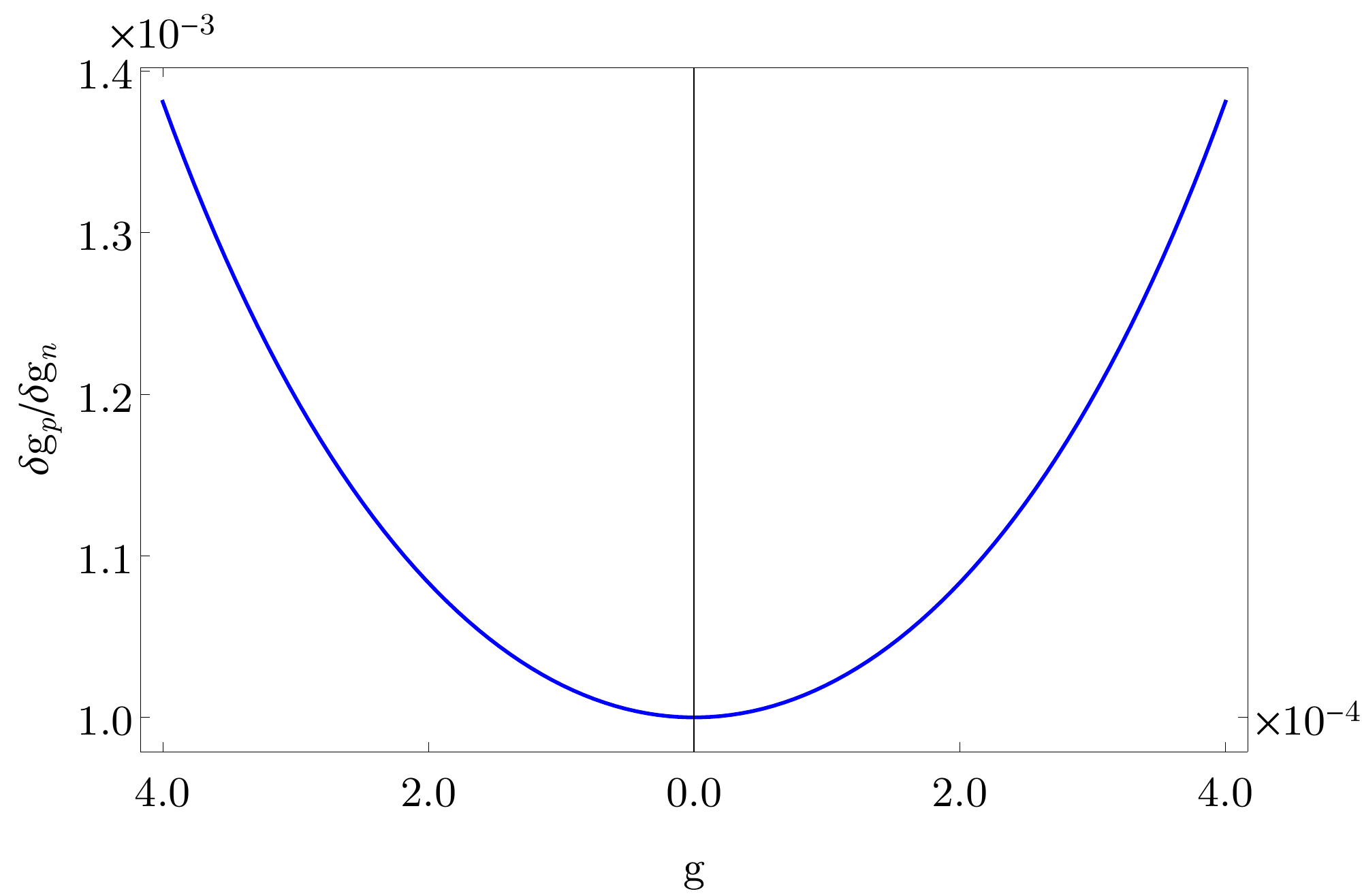}
\caption{Ratio between the systematic error with and without postselection
of the system as a function of the interaction parameter $g$.}
\label{fig:lp1gOverlp3gPlot}
\end{figure}

Fig. \ref{fig:lp1gOverlp3gPlot} shows the ratio $\frac{\dg[p]}{\dg[n]}$
for a range of values of the interaction parameter $g$. The figure
shows that the ratio varies approximately parabolically with $g$.
This is due to the second order terms of $g$ which are neglected
in the weak value formalism and our results. In spite of this, the
figure still shows that the suppression ratio of the systematic error
has the order $10^{-3}$, approximately the inverse of the weak value
$1/\delta$, which matches the theoretical results we obtained.

\begin{comment}
Fig. \ref{fig:lp1sOverlp3sPlot} shows the ratio $\frac{\dg[p]}{\dg[n]}$
with different system decoherence strength $\es$, in order to consider
the influence of $\es$ on the weak value suppression of the systematic
error. Since the systematic error without postselection $\dg[n]$
does not depend on $\es$, whereas the systematic error with postselection
$\dg[p]$ does, we can expect from our previous theoretical results
to see suppression of the systematic error. Indeed, the figure shows
how the postselection contributes to such error suppression. Furthermore,
we also see that as the absolute value of the system decoherence strength
becomes larger, the deviations from the first-order approximation
in the weak value formalism also become increasingly more noticeable.

% \begin{figure}[!htbp]% \includegraphics[scale=0.42]{figures/lp1sOverlp3sPlot.pdf} \caption{Ratio between the systematic error with and without postselection% as a function of $\epsilon_{S}$.}% \label{fig:lp1sOverlp3sPlot}% \end{figure}
\end{comment}

\begin{figure}[!htbp]
\includegraphics[width=0.95\linewidth]{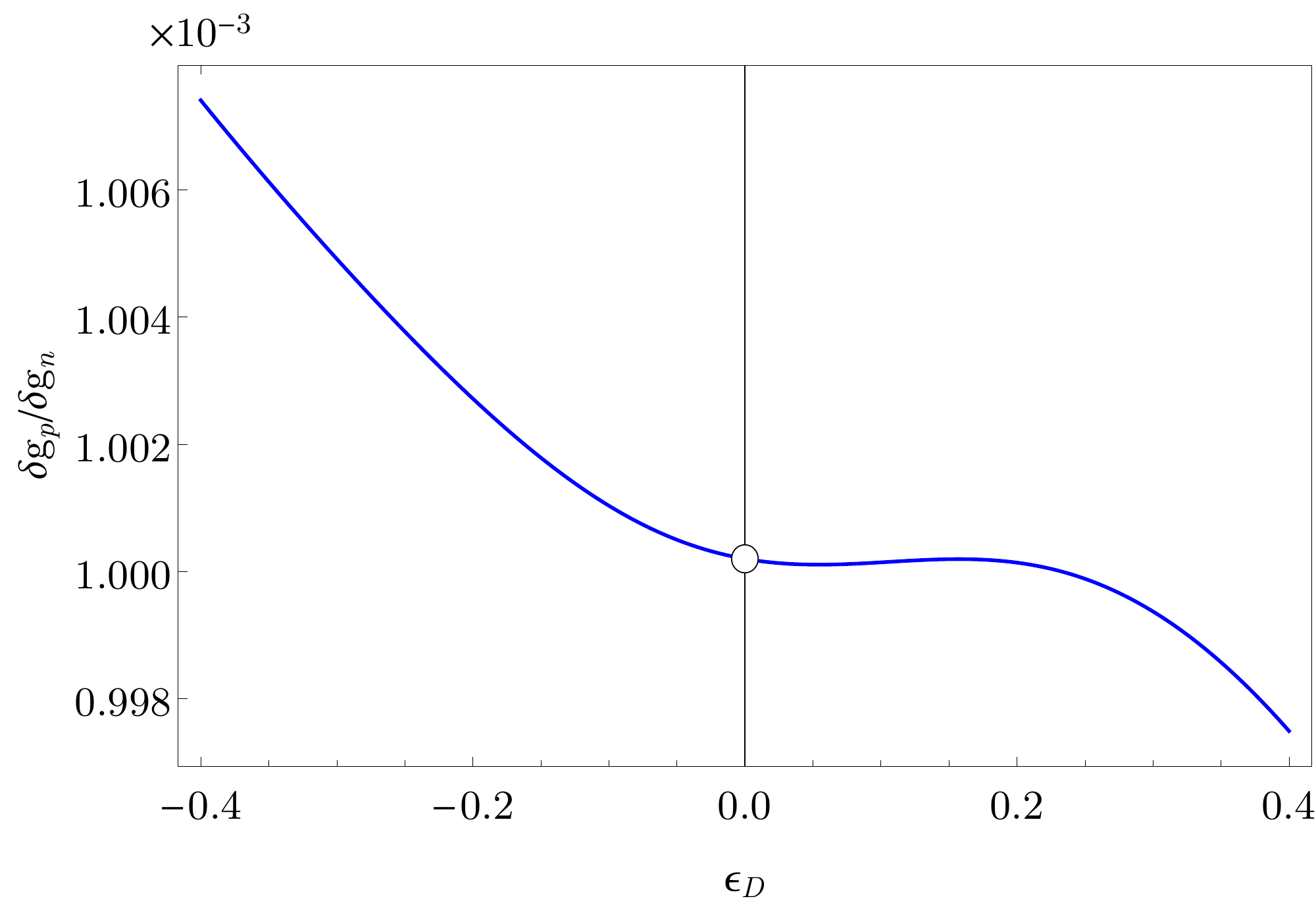}
\caption{Ratio between the systematic error with and without postselection
of the system as a function of the probe decoherence strength $\protect\ed$.
$\protect\ed=0$ is removed from the plot, since the systematic errors
for both postselected and standard weak measurements are zero at that
point, leading to an ill-defined ratio between them.}
\label{fig:lp1pOverlp3pPlot}
\end{figure}

Fig. \ref{fig:lp1pOverlp3pPlot} shows the ratio $\frac{\dg[p]}{\dg[n]}$
with different probe decoherence strength $\ed$ in order to consider
the effect of weak value amplification on suppressing systematic errors
for different $\ed$. It can be seen from the figure that the suppression
rate of systematic error changes very little around $10^{-3}$ for
a wide range of $\ed$, which implies that the suppression of systematic
error by weak value amplification is very stable with respect to the
strength of the probe decoherence.

\begin{comment}
Fig. (2a) and (2b) showed that as $\es$ increases, the systematic
error $\dg[all]$ also increases, yet it is still suppressed by the
order of $\delta$. A more interesting result is Fig. (2c). When $\es$
increases from $0$ to $10^{-2}$, ${\displaystyle \frac{\dg[p]}{\dg[all]}}$
increases from smaller than $1$ to larger than $1$. This is because
$\es$ gives the same order of contribution to $\dg[all]$ as $\ed$
does according to (\ref{eq:12}). And if the contribution from $\es$
is opposite to that from $\ed$, the systematic error from $\ed$
can be partially canceled, which is what (\ref{eq:12}) implied.
\end{comment}

\section{Discussion}

In this paper, we mainly considered the systematic error on weak measurement
caused by decoherence, and studied the advantage of weak value amplification
in suppressing the systematic error of the measurements. We find that
a large weak value can effectively suppress the systematic error of
postselected weak measurements, compared to standard weak measurement.
This is distinct from the loss of Fisher information in the postselected
weak measurement under decoherence of the system \cite{Knee2013}.

Aside from postselecting the system to one specific state and discarding
the unselected results (which was studied in this paper), an alternative
method is to retain all the postselection results and use them to
estimate the interaction parameter in the Hamiltonian. It has been
proven that this method can retain more Fisher information than the
method discarding failed postselection results \cite{Tanaka2013,Combes2014,Ferrie2014,Knee2014,Zhang2015a}.
This is easy to understand, because the total Fisher information of
the data for the estimation is proportional to the size of the data
according to the Cramér-Rao bound (\ref{eq:camer-rao-bound}), although
in all experimental cases to date, the amount of extra information
gained by retaining the other postselection outcomes is negligible
\cite{Hosten2008,Dixon2009,Starling2009,Viza2015a}.

However, the question of whether retaining all postselection results
can improve the systematic error in this scenario is more complex.
On the one hand, the systematic error is not proportional to the size
of data (see Eq. \eqref{eq:camer-rao-bound}), and the Fisher information
in the first-order solution to the systematic error \eqref{eq:sys-error}
is the average Fisher information of a single event. Therefore, retaining
all postselection results, which mixes the events that have high Fisher
information and that have low Fisher information, may lead to a lower
average Fisher information than postselecting the system, which selects
the high Fisher information events only. Thus the systematic error
may increase. On the other hand, retaining all postselection results
averages the relative entropy between the ideal and the real probability
distributions, and the low postselection probabilities for those high
Fisher information events could lower the average relative entropy,
which may compensate for the loss in average Fisher information. Therefore,
it is not clear whether retaining the failed postselection results
can in principle improve the systematic error of weak measurement
as it does for the Fisher information.

We leave this problem as an open question for future research.
\begin{acknowledgments}
The authors thank Justin Dressel for helpful discussions. SP and ANJ
acknowledge the support from the US Army Research Office under Grants
No. W911NF- 15-1-0496 and No. W911NF-13-1-0402 and the support from
the National Science Foundation under Grant No. DMR- 1506081. SP,
JRGA, and TAB also thank the support from the ARO MURI under Grant
No. W911NF-11-1-0268.
\end{acknowledgments}

\bibliographystyle{apsrev4-1}
\bibliography{WVA_docoherence}

\end{document}